\documentclass{article}[12pt]
\usepackage{graphicx}
\usepackage{hyperref}
\usepackage{amssymb}
\usepackage{multirow}
\usepackage{caption}
\usepackage{subcaption}
\usepackage[left=1.5cm,right=1.5cm,top=1.5cm,bottom=1.5cm]{geometry}
\usepackage{authblk}
\usepackage{amsmath}
\begin{document}
	\title{The role of finite value of strange quark mass $(m_{s}\neq0)$ and baryon number density $(n)$ on the stability and maximum mass of strange stars}
	\author[1]{Pradip Kumar Chattopadhyay\thanks{pkc$_{-}76$@rediffmail.com}}
	\author[2]{Debadri Bhattacharjee\thanks{debadriwork@gmail.com}}
	\affil[1,2]{IUCAA Centre for Astronomy Research and Development (ICARD), Department of Physics, Cooch Behar Panchanan Barma University, Vivekananda Street, District: Cooch Behar, \\ Pin: 736101, West Bengal, India}
	\maketitle
\begin{abstract}
	This study describes the impact of non-zero value of strange quark mass $(m_{s})$ and number density of baryons $(n)$ on the structure, stability and maximum mass of strange stars. We derive an exact relativistic solution of the Einstein field equation using the Tolman-IV metric potential and modified MIT bag model EoS, $p_{r}=\frac{1}{3}(\rho-4B')$, where $B'$ is a function of bag constant $B$, $m_{s}$ and baryon number density $(n)$. Following CERN's findings, transition of phase from hadronic matter to Quark-Gluon Plasma (QGP) may occur at high densities in presence of favourable conditions. The standard MIT bag model, with a constant $B$, fails to explain such transition properly. Introducing a finite $m_{s}$ and Wood-Saxon parametrisation for $B$, dependent on baryon number density $(n)$, provides a more realistic EoS to address such phase transition. Both $m_{s}$ and $n$ constrain the EoS, making it softer as $m_{s}$ increases. Solutions to the TOV equations reveal that for massless strange quarks, maximum mass is 2.01 $M_{\odot}$ and corresponding radius is 10.96 Km when $n=0.66~fm^{-3}$. These values decrease to 1.99 $M_{\odot}$ and 1.96 $M_{\odot}$, with corresponding radii of 10.88 Km and 10.69 Km for $m_{s}=50$ and $100~MeV$ respectively having same $n$ value. It is interesting to note that a corelation exists between $n$ and $m_{s}$. The hadronic to quark matter transition occurs at higher values of $n$, when $m_{s}$ increases such as $n\geq0.484,~0.489$ and $0.51~fm^{-3}$ for $m_{s}=50$ and $100~MeV$ respectively. Beyond these values, the energy per baryon $(\mathcal{E_{B}})$ drops below $930.4~MeV$, indicating a complete transition to quark matter. For physical analysis, we have considered $n~(=0.578~fm^{-3})$ which lies in the stable region with $B(n)=70~MeV/fm^{3}$. The model provides a viable description of strange stars, satisfying all necessary physical requirements.
\end{abstract}
\section{Introduction}\label{sec1}
In recent decades, compact objects have become a focal point of research within the framework of relativistic astrophysics. Current research, encompassing both the theoretical as well as experimental methodologies, is focused on understanding the composition of the interior matter of compact objects subjected to extreme densities as well as developing the equation of state (EoS), which defines the relationship between pressure and density, under these conditions. This understanding is essential for accurately modeling compact objects. The final evolutionary stage of stars, compact objects, are categorized into different subclasses based on the mass of their stellar remnants \cite{Shapiro}. Among these, Neutron Stars (NS) are most fascinating objects. Baade and Zwicky proposed the concept of NSs \cite{Baade}, suggesting that they form as a result of supernova explosions. Their small size and extremely dense cores ($\sim2.7\times10^{14}$ $g/cm^{3}$ or higher) make NSs ideal candidates for studying extreme physical conditions. Oppenheimer and Volkoff \cite{Oppenheimer} were the first to calculate a solution for NSs by modeling neutrons as ideal Fermi gases, estimating a mass limit of $0.7~M_{\odot}$, where $M_{\odot}$ is the solar mass. However, at such high-density regimes, interactions between nucleons must be considered, making the maximum mass of NSs highly dependent on the chosen EoS. Despite extensive research, an exact definitive EoS remains elusive. Interestingly, some observed compact objects exhibit smaller masses and radii compared to standard NSs but possess greater compactness (higher mass-to-radius ratio). These characteristics cannot be explained by conventional NS models, necessitating new approaches to understand their observed properties. In this context, the hypothesis of quark stars, composed of Strange Quark Matter (SQM), has gained widespread acceptance. According to Bodmer \cite{Bodmer} and Witten \cite{Witten}, it was proposed that, under the extreme conditions, ground state of matter may consist of deconfined quark phases rather than the most stable nuclei, $^{56}Fe$. The key argument for this theory is that SQM has a lower energy per baryon compared to $^{56}Fe$, suggesting a comparatively stable structure. Itoh \cite{Itoh} first considered quark stars in hydrostatic equilibrium, and Madsen \cite{Madsen} demonstrated that, under zero external pressure, the inclusion of strange quarks $(s)$ alongside up $(u)$ and down $(d)$ quarks is necessary to achieve an energy per baryon of lower value. This theoretical framework has led to the classification of a new family of compact objects known as Strange Stars (SS) \cite{Madsen, Baym, Glendenning, Alcock1}. Although the theoretical foundations of SSs are well-established, the exact process of their formation remains unclear. Two primary formation mechanisms have been proposed \cite{Bodmer, Witten, Itoh}: (i) a phase transition from hadron to quark occurring at extremely high temperatures in the interior of compact objects in the early universe, and (ii) the possibility that a neutron star may undergo spontaneous compression to form a SS due to ultra-high densities in presence of strong gravitational forces. Alford \cite{Alford} further suggested that in the core of a NS, low temperatures and extreme densities could be sufficient to create quark matter in bulk.

To explain the relevant properties of SSs, selecting an appropriate EoS is crucial. In this regard, the MIT bag model EoS \cite{Kapusta, Chodos1} has emerged as a widely accepted framework. This model assumes that quarks form Fermi gas of degenerate nature consisting of $u,~d$ and $s$ quarks, along with electrons $e^{-}$, and the EoS is formulated as:
\begin{equation}
	p=\frac{1}{3}(\rho-4B) \label{eq1}
\end{equation}
In this model, $p$ represents the pressure, $\rho$ denotes the energy density of matter and the term $B$ is known as bag constant. Within the framework of the MIT bag model, Farhi and Jaffe \cite{Farhi} investigated the influence of the strange quark mass $(m_{s})$ on the overall properties of SQM. Their research identified a stability region for SQM, expressed as a "stability window", in the $(m_{s}-B)$ parameter space. Extensive studies have been performed to investigate the characteristics of SS using the MIT bag model \cite{Brilenkov,Paulucci,Arbanil,Lugones,Chowdhury,Maharaj,Abbas,KBG,KBG1,KBG2}.

To achieve a more precise physical representation, it is essential to consider that in case of a highly dense compact object, interior fluid may have an anisotropic nature. Ruderman \cite{Ruderman} and Canuto \cite{Canuto} suggested that interior of a compact star might develope anisotropic pressure when their matter density surpasses the density of nuclear matter . Anisotropy in compact star cores can be explained by type 3A superfluidity, which may be a significant factor \cite{Kippenhahn}. From a quantum mechanical standpoint, type 3A superfluidity is a captivating macroscopic theory. According to Pauli's exclusion principle, nucleons cannot occupy the same energy state as they are fermions. However, at extremely short distances, strong repulsion between nucleons occurs, resisting gravitational collapse in NS. At low temperatures, Cooper pairs \cite{Broglia} may form by nucleons, which behave like bosons, altering the situation. Thus, at very low temperatures and on large scales, nucleons exhibit collective behaviour similar to a non-viscous nucleonic condensate, akin to He-3 superfluids. In the interior of NS, the presence of high pressure significantly increases the matter temperature to billions of degrees raising the critical temperature required for the formation of Cooper pairs and enabling superfluidity into matter under extreme conditions. Three types of superfluids may exist within the core of NS \cite{Page}. The anisotropic behaviour of matter can be linked to electromagnetic fields and fermionic fields in NS \cite{Sawyer}, pion condensation \cite{Sawyer1}, viscosity \cite{Herrera}, and more. Bowers and Liang \cite{Bowers} introduced pressure anisotropy in their study of compact stars to examine stellar characteristics such as compactness, mass-radius relationship, surface redshift, stability etc. Studies on the relativistic properties of different anisotropic NS suggest that if one considers higher and higher values of anisotropic factor, there is practically no theoretical upper limit of maximum mass and surface redshift as proposed by Heintzmann and Hillebrandt \cite{Heintzmann}. Extensive studies have been conducted on spherically symmetric stellar configurations composed of anisotropic matter and their static equilibrium \cite{Maurya,Maurya1,Deb,Kalam,Mak,Mak1,Hernandez}, following the discoveries of Carter and Langlois \cite{Carter}. Recently, Bhattacharjee and Chattopadhyay \cite{Bhattacharjee} examined the effect of Chaplygin EoS in modified form on anisotropic stellar structures. Anisotropic generalisation of spherically symmetric fluid sphere provides a more universal approach to utilising the EoS. Tolman-Oppenheimer-Volkoff equation \cite{Tolman,Oppenheimer} can uniquely be applied to determine the physical parameters of some compact stars, including their potential mass and radius relation using known EoS. 

Tolman \cite{Tolman} explored a range of solutions to the Einstein Field Equations (EFE) for perfect fluids, generating eight solutions by linking the metric functions through a single equation and imposing specific constraints. Among these, Tolman-IV offers a straightforward form, accommodating a sphere of compressible fluid with nonzero central density and pressures. Therefore, to obtain a tractable exact solution of the EFE in this present formalism, we have considered the Tolman-IV metric ansatz \cite{Tolman}. Building on the Tolman-IV potential, Banerjee \cite{Banerjee} derived a well-behaved and regular solution of EFE within the framework of MIT bag model EoS, allowing for anisotropic matter distribution. Bhar et al. \cite{Bhar} developed a compact stellar model using the Tolman IV potential in an anisotropic context, demonstrating its regularity and stability. Additionally, Das et al. \cite{Das} proposed a specific form of the $g_{rr}$ metric component, which can be seen as an extension of the Tolman IV potential for anisotropic matter. Several articles \cite{Maharaj1,Takisa,Takisa1,Thirukkanesh,Thirukkanesh1,Rocha} extensively examined the properties of Strange Quark Stars (SQS), contributing to our understanding of these phenomena. 

In the MIT bag model, as expressed in Eq.~(\ref{eq1}), the constant $B$ represents the pressure at the bag wall that confines quarks. Accurate simulation of EoS of the interior matter in question is note possible if one neglects the possible interaction between the surrounding medium of high density and the quarks. Thus, understanding the interior of compact stars made up with quark matter remains challenging. The MIT bag model assumes that quarks are confined within a perturbative vacuum or "bag," resulting from the net inward pressure $B$. The pressure is exerted by the surrounding medium of nonperturbative vacuum. As the baryon number density $n$ increases, the difference between these perturbative and non-perturbative vacua may vanish, requiring $B$, the net inward pressure, $B$, to also vanish. It is thus more appropriate to consider a density-dependent $B$ \cite{Reinhardt} or baryon number-dependent \cite{Burgio} $B$. Moreover, the introduction of finite strange quark mass $(m_{s}\neq0)$ is very relevant in realistic SS modeling since the presence of $m_{s}$ significantly challenges the conventional SS models and it influences the structural and dynamical properties of SS. As finite value of strange quark mass $(m_{s}\neq0)$ and baryon number density have significant effect on the energy per baryon $(\mathcal{E_{B}})$ of system, our aim in adopting the nonzero strange quark mass $(m_{s})$ and baryon number density dependent $B$ is to develop a more realistic EoS for predicting key properties of SS. These properties include mass-radius relations, stability criteria and causality conditions as well as validity of essential energy conditions. We utilise parametrisation of $B$ similar to a form proposed by Wood-Saxon \cite{Wood}, where $B$ varies with baryon number density $(n)$.

The rest of the paper is arranged in the following manner: in Sect.~\ref{sec2}, we have discussed the thermodynamics of SQM at absolute zero temperature and the baryon number density dependent bag model. In this section, we have plotted the variation of energy per baryon $(\mathcal{E_{B}})$ against the baryon number density $(n)$ and strange quark mass $(m_{s})$ to determine the permissible restrictions on both which ultimately lead to the particular choice of bag constant $(B)$ in this model. Sect.~\ref{sec3} addresses the exact relativistic solution of EFE considering the Tolman-IV metric ansatz and the modified MIT bag model EoS. We address the necessary boundary conditions by matching the interior solutions with the exterior vacuum Schwarzschild solution in Sect.~\ref{sec4}. In Sect.~\ref{sec5}, we have numerically solved the TOV equations to determine the possible maximum mass and the associated radius in the present model. Moreover, we have also predicted the radii of a wide class of compact stars in this section. The physical viability of the present model, in terms of the radial variations of characteristic parameters along with the causality and necessary energy conditions, is demonstrated in Sect.~\ref{sec6}. The stability analysis is described in Sect.~\ref{sec7}. Finally, we summarise the main findings of the present study in Sect.~\ref{sec8}.   
\section{Thermodynamics of SQM at $T\rightarrow0$ and $n$-dependent bag model} \label{sec2} 
The quark matter in the interior of a SS of high density is described by the EoS as proposed by MIT bag model \cite{Chodos}. Stellar interior can be modeled as a degenerate Fermi gas of quarks, as quarks are fermions, in which the quarks exist in a de-confined phase. If $A$ represents the total baryon number present inside a star, the corresponding deconfined quark matter consists of $3A$ quarks, forming a colour-neutral singlet state of Fermi gas. According to the MIT bag model, the behaviour of quark confinement described in Refs.~\cite{Chodos, Kettner} is approximated as:
\begin{equation}
	p+B=\sum_{i=u,d,s,e^{-}}p_{i}, \label{eq2}
\end{equation}   
and 
\begin{equation}
	\rho-B=\sum_{i=u,d,s,e^{-}}\rho_{i}, \label{eq3}
\end{equation}
where, $p_{i}$, $\rho_{i}$ and $B$ respectively denote the pressure and energy density associated with the $i^{th}$ particle and bag constant. In this context, we assume that strange matter consists of three types of quarks, namely, up $(u)$, down $(d)$, and strange $(s)$, along with electrons $(e^{-})$. According to Kettner et al. \cite{Kettner}, the potential existence of charm quark stars is ruled out due to their instability against radial oscillations. Additionally, since the density of muons only becomes significant beyond a certain threshold of density of charm quark, we can disregard the presence of muons \cite{Kettner}. In SQS, Coulomb interactions dominate over gravitational effects, resulting in SQM being approximately charge neutral at the lowest possible energy state \cite{Alcock}. The condition for charge neutrality is expressed as:
\begin{equation}
	\sum_{i=u,d,s,e^{-}}n_{i}q_{i}=0, \label{eq4}
\end{equation}
where, symbol $i$ indicates the $i^{th}$ particle. If the chemical potential $(\mu)$ of quarks exceeds the absolute temperature of the stellar configuration, the pressure $(p_i)$, energy density $(\rho_i)$ of quarks as well as number density $(n_i)$ can be approximated following the approach given in \cite{Kettner, Peng} and are given below in the limit $T \rightarrow 0$:
\begin{eqnarray}
	p_{i}=\frac{g_{i}\mu_{i}^{4}}{24\pi^{2}}\sqrt{1-\Big({\frac{m_{i}}{\mu_{i}}}\Big)^{2}}\Big\{{1-\frac{5}{2}\Big(\frac{m_{i}}{\mu_{i}}\Big)^{2}}\Big\} \nonumber\\+\frac{3}{2}\Big(\frac{m_{i}}{\mu_{i}}\Big)^{4}\ln{\frac{1+\sqrt{1-\Big(\frac{m_{i}}{\mu_{i}}\Big)^{2}}}{\Big(\frac{m_{i}}{\mu_{i}}\Big)}}, \label{eq5}
\end{eqnarray} 
\begin{eqnarray}
	\rho_{i}=\frac{g_{i}\mu_{i}^{4}}{24\pi^{2}}\sqrt{1-\Big({\frac{m_{i}}{\mu_{i}}}\Big)^{2}}\Big\{{1-\frac{1}{2}\Big(\frac{m_{i}}{\mu_{i}}\Big)^{2}}\Big\}\nonumber\\-\frac{1}{2}\Big(\frac{m_{i}}{\mu_{i}}\Big)^{4}\ln{\frac{1+\sqrt{1-\Big(\frac{m_{i}}{\mu_{i}}\Big)^{2}}}{\Big(\frac{m_{i}}{\mu_{i}}\Big)}}, \label{eq6}
\end{eqnarray}
\begin{equation}
	n_{i}=\frac{g_{i}\mu_{i}^{3}}{6\pi^{2}}\Big[1-\Big(\frac{m_{i}}{\mu_{i}}\Big)^{2}\Big]^{\frac{3}{2}}, \label{eq7}
\end{equation}
where, $g_{i}$ represents degeneracy factor. The degeneracy factors for the quarks and the electrons are $g_{i}=6$ and $g_{i}=2$ respectively. Hence, the condition for charge neutrality as expressed in Eq.~(\ref{eq4}) is modified to the following form:
\begin{equation}
	2\Big(1-\frac{\mu_{e^{-}}}{\mu}\Big)^{3}-\Big(\frac{\mu_{e^{-}}}{\mu}\Big)^{3}-\Big\{1-\Big(\frac{m_{i}}{\mu_{i}}\Big)^{2}\Big\}^{\frac{3}{2}}-1=0, \label{eq8}
\end{equation}
Here, $\mu = \mu_d = \mu_s$ represents chemical potential of $d$ and $s$ quarks. The $u$ and $d$ quarks are significantly lighter compared to the $s$ quarks and when $m_s \rightarrow 0$ is applied in the modified charge neutrality condition, Eq.~(\ref{eq8}), we find that $\mu_{e^{-}} \rightarrow 0$, implying that electrons are not crucial for maintaining charge neutrality in a system of massless quarks. Under these conditions, the equation of state (EoS) $p_i = \frac{1}{3} \rho_i$ can be derived from Eqs.(\ref{eq4}) and (\ref{eq5}). By utilizing Eqs.(\ref{eq2}) and (\ref{eq3}), the MIT bag model EoS for a system of massless strange quarks can be constructed as \cite{Kapusta}:
\begin{equation}
	p_{r}=\frac{1}{3}(\rho-4B). \label{eq9}
\end{equation}
Incorporating a finite value strange quark mass $(m_{s}\neq0)$ into our analysis, alongside Eqs.~(\ref{eq2}), (\ref{eq3}), (\ref{eq5}), and (\ref{eq6}), yields the most general form of the EoS for SQM, which can be expressed as follows:
\begin{equation}
	\rho=3p_{r}+4B+\rho_{s}-3p_{s}, \label{eq10} 
\end{equation}
\begin{equation}
	\Rightarrow p_{r}=\frac{1}{3}(\rho-B'), \label{eq11}
\end{equation}
where, $B'=B+\frac{1}{4}(\rho_{s}-3p_{s})$ and the density of strange quarks is denoted by $\rho_s$, and the pressure resulting from the presence of strange quarks is $p_s$. 

\begin{figure}[h]
	\centering
	\includegraphics[width=8cm]{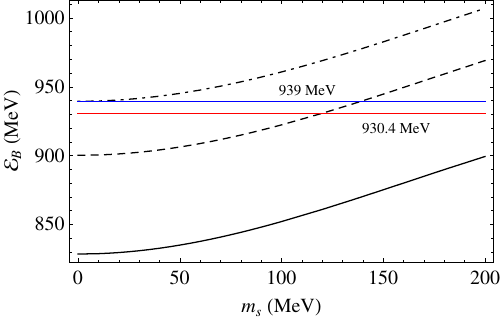}
	\caption{$\mathcal{E_{B}}$ vs. $m_{s}$ for different choices of $B(n)$. Here, solid, dashed and dot-dashed lines correspond to $B(n)=57.55,~80.21$ and $95.11~MeV/fm^{3}$ respectively.}
	\label{fig1}
\end{figure}

\begin{figure}[h]
	\centering
	\includegraphics[width=8cm]{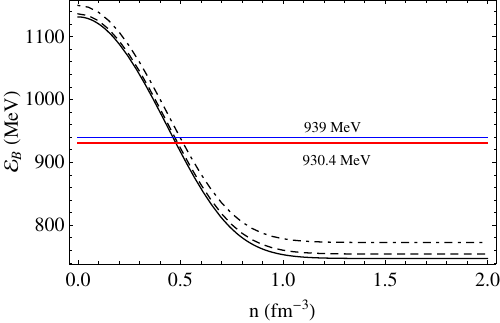}
	\caption{Graphical variation of $\mathcal{E_{B}}$ with $n$. Here, solid, dashed and dot-dashed lines correspond to $m_{s}=0,~50$ and $100~MeV$ respectively.}
	\label{fig2}
\end{figure}

In the high density regime, an ultimate phase transition from the hadronic matter to Quark Gluon Plasma (QGP) is anticipated to occur \cite{Witten,Baym,Glendenning}. However, the MIT bag model may be employed to describe the de-confined phase of quarks \cite{Chodos,Kapusta}. The CERN experiments have revealed that the QGP formation in the heavy-ion collisions is characterised by the small baryon number density and high temperature. Whereas, the quark phase illustrates high baryon number density and low temperature. In this framework, when the MIT bag model is applied, the de-confined quark phase emerges at a constant quark-gluon energy density regardless of the thermodynamic aspects of the model \cite{Cleymans}. By calibrating the parameters of the MIT bag model to align with CERN data, we can restrict our analysis to zero temperature, considering that the energy density remains constant along the phase transition line. Following the work of Burgio et al. \cite{Burgio}, we observe that along the transition line from hadronic matter to the quark phase, the energy densities become equal at a specific point. CERN-SPS results indicate that this intersection point corresponds to an energy density of $1.1~GeV/fm^3$. In our current model, we put forward that the hadron-quark phase transition is solely dependent on the energy density, and the QCD phase diagram in the temperature and chemical potential plane is characterised exclusively by the bag constant $B$. To characterise this phase transition in relation to CERN experiment results and to apply the quark core hypothesis in stellar modeling, we adopt an extreme Wood-Saxon-like characterisation \cite{Wood} for the baryon number-dependent bag parameter $B(n)$ and estimate the value of $B$ as \cite{Burgio}:
\begin{equation}
	B(n)=B_{as}+(B_{0}-B_{as})e^{-\beta_{n}(\frac{n}{n_{0}})^2}, \label{eq12}
\end{equation}
where, $B_{as}=38~MeV/fm^{3}$ represents the value of $B$ at asymptotic density, $B_{0}=B(\rho=0)=200~MeV/fm^{3}$, $\beta_{n}=0.14$ is a free characteristic parameter arising from calibration of the densities of nuclear and quark matter at $1.1~GeV/fm^3$, $n$ is the baryon number density and $n_{0}=0.17~fm^{-3}$ is the baryon number density of the ordinary matter sector. Hence, in the present formalism, the modified bag constant takes the form:
\begin{equation}
	B'=B(n)+\frac{1}{4}(\rho_{s}-3p_{s}). \label{eq12a}
\end{equation}  
Another important aspect in the SS modeling is the determination of energy per baryon $(\mathcal{E_{B}})$. The energy per baryon $(\mathcal{E_{B}})$ for the most stable nucleus, $^{56}Fe$, is $930.4MeV$. For quark matter consisting of two-flavour quarks $(u$ and $d)$, the energy per baryon would exceed this value. Otherwise, $^{56}Fe$ would be composed of two-flavour quarks, $u$ and $d$, which has not been observed experimentally in nature. Inclusion of strange quark $(s)$ into the system consisting of $u$ and $d$ quarks effectively decreases the energy per baryon. Therefore, a three-flavour quark system is considered to be stable when its energy per baryon is less than $930.4~MeV$. The metastability condition imposes a constraint, keeping the energy per baryon within the range of $940.4~MeV < \mathcal{E_{B}} < 939MeV$ \cite{Backes}, where the upper limit aligns with the typical nucleon mass range. When $\mathcal{E_{B}} > 939~MeV$, SQM becomes unstable. The baryon number density for a three-flavour SQM system is expressed as follows:
\begin{equation}
	n=\frac{1}{3}\sum_{i=u,d,s}n_{i}. \label{eq13}
\end{equation} 
Using Eq.~(\ref{eq12}), we have tabulated the numerical values of baryon number density $(n)$ for different choices of bag constant $B(n)$ in Table~\ref{tab1}. The values of $B(n)$ are restricted within the range of $57.55~MeV/fm^{3}<B(n)<95.11~MeV/fm^{3}$ necessary for stable SQM at zero external pressure \cite{Madsen}.  
\begin{table}[h]
	\centering
	\caption{Determination of baryon number density $(n)$ for different choices of bag parameter $(B(n))$}
	\label{tab1}
	\begin{tabular}{cc}
		\hline
		Bag constant $(B(n))$ & Baryon number density $(n)$ \\
		$(MeV/fm^{3})$ & $(fm^{-3})$ \\ \hline
		57.55 & 0.660 \\
		70 & 0.578 \\
		75 & 0.552 \\
		80.21 & 0.526 \\
		95.11 & 0.464 \\
		\hline
	\end{tabular}
\end{table} 
We have constrained the numerical choice of strange quark mass $(m_{s})$ through the depiction of $(\mathcal{E_{B}})$ vs. $m_{s}$ plane in Fig.~(\ref{fig1}) for different choices of bag constant $B(n)$ as tabulated in Table~\ref{tab1}. 
From Fig.~(\ref{fig1}), it is evident that there exists some bounds on the choice of $m_{s}$ that corresponds to the stable, metastable and unstable SQM. In Table~\ref{tab2}, we tabulate the permissible range of $m_{s}$ for the choices of $B(n)$ considered here. 
\begin{table}[ht!]
	\centering
	\caption{Stability window of 3 flavour SQM with constrained $m_{s}$ and different bag value $(B)$.} 
	\label{tab2}
	\begin{tabular}{cccc}
		\hline
		\multirow{2}{*}{ Stability }& $B(n)=57.55$ & $B(n)=80.21$ & $B(n)=95.11$ \\ 
		& $(MeV/fm^{3})$ & $(MeV/fm^{3})$ & $(MeV/fm^{3})$ \\ \hline
		Stable & \multirow{2}{*}{$m_{s}<290~MeV$} & \multirow{2}{*}{$m_{s}<119~MeV$} & \multirow{2}{*}{....}\\ 
		$(\mathcal{E_{B}}<930.4~MeV)$ &&& \\ \hline
		MetaStable & \multirow{2}{*}{....} & \multirow{2}{*}{$119<m_{s}<138~MeV$} & \multirow{2}{*}{....}\\
		$(930.4<\mathcal{E_{B}}<939~MeV)$ &&& \\\hline
		Unstable & \multirow{2}{*}{....}  & \multirow{2}{*}{$m_{s}>138~MeV$} & \multirow{2}{*}{$m_{s}>0~MeV$}\\
		$(\mathcal{E_{B}}>939~MeV)$ &&& \\ \hline
	\end{tabular}
\end{table} 
Within the defined stability range, we have selected $B(n) = 70~MeV/fm^{3}$ to generate a stable SS model. Moreover, we have noted some intriguing results from the variation of energy per baryon $(\mathcal{E_{B}})$ with respect to the baryon number density $(n)$ as shown in Fig.~(\ref{fig2}). From Fig.~(\ref{fig1}), it is observed that energy per baryon $(\mathcal{E_{B}})$ increases with the increase of $m_{s}$ whereas from Fig.~(\ref{fig2}), it is evident that $\mathcal{E_{B}}$ decreases with increasing baryon number density $(n)$. Thus, a corelation exists between $m_{s}$ and $n$, so far the stability of the SQM is concerned.

Fig.~(\ref{fig2}) demonstrates the impact of strange quark mass $(m_{s})$ on the $\mathcal{E_{B}}$ vs. $n$ diagram. It is interesting to note that the matter shows unstable behaviour in relation to quark phase when $n<0.464~fm^{-3}$, $n<0.473~fm^{-3}$ and $n<0.50~fm^{-3}$, for $m_{s}=0,~50$ and $100~MeV$, respectively, which in turn reflects the fact that the matter exhibits a stable hadronic nature. Thus, if we consider higher values of $m_{s}$, hadronic phase exists for higher values of $n$. Furthermore, a probable transition from the hadron to quark phase is anticipated for $n\geq0.464,~0.473$ and $0.50~fm^{-3}$ respectively for $m_{s}=0,~50$ and $100~MeV$. Presumably, there may be a coexistence of hadrons and quarks in this situation which is referred to as a metastable state. The metastability is ensured for $0.464\leq n\leq0.482~fm^{-3}$, $0.473\leq n\leq0.489~fm^{-3}$ and $0.50\leq n\leq0.51~fm^{-3}$ for $m_{s}=0,~50$ and $100~MeV$ respectively. It is evident that the range of baryon number density $(n)$, for which metastable window exists, is reducing with the increase of $m_{s}$. Eventually, with increasing baryon number density, i.e., $n\geq0.482,~0.489$ and $0.51~fm^{-3}$, for $m_{s}=0,~50$ and $100~MeV$ respectively, the energy per baryon attains a value $<930.4~MeV$, which implies the conversion of the entire hadron assembly into quarks and the matter transforms into a stable state. Additionally, from Table~\ref{tab1}, we note that the choice of $B(n) = 70~MeV/fm^{3}$ provides $n=0.578~fm^{-3}$ which abides by the prescribed range of $n$ for different choices of $m_{s}$.       
\section{Solution of Einstein field equations} \label{sec3}
In the static and spherical curvature co-ordinate system (t,r,$\theta$, $\phi$), the line element takes the form:
\begin{equation}
	ds^2=-e^{2\nu(r)}dt^2+e^{2\lambda(r)}dr^2+r^2(d\theta^2+sin^2\theta d\phi^2). \label{eq14}
\end{equation}
In relativistic units, $G=1$ and $c=1$, the most general form of the energy-momentum tensor $T_{ij}$ in presence of pressure anisotropy is given by 
\begin{equation}
	T_{ij}=diag(-\rho,p_{r},p_{t},p_{t}). \label{eq15}
\end{equation}
In the relativistic framework, the Einstein's field equations (EFE) associating the 4 dimensional space-time geometry and the matter content are expressed as:
\begin{equation}
	R_{ij}-\frac{1}{2}g_{ij}R=8\pi T_{ij}, \label{eq16}
\end{equation}

Employing Eqs.~(\ref{eq14}) and (\ref{eq15}) in Eq.~(\ref{eq16}), we get the following tractable set of the EFEs in the following form:
\begin{eqnarray}
	\frac{2e^{-2\lambda}\lambda'}{r}+\frac{(1-e^{-2\lambda})}{r^2}=8\pi\rho, \label{eq17} \\
	\frac{2e^{-2\lambda}\nu'}{r}-\frac{(1-e^{-2\lambda})}{r^2}=8\pi p_{r}, \label{eq18} \\
	e^{-2\lambda}(\nu''+\nu'^2-\lambda'\nu'+\frac{\nu'}{r}-\frac{\lambda'}{r})=8\pi p_{t},\label{eq19}
\end{eqnarray}
where, prime $(')$ denotes the derivatives with respect to $r$. Moreover, the pressure anisotropy $(\Delta)$ is described as the difference between the radial and tangential components of pressure, ${\it viz.}$, 
\begin{equation}
	\Delta=p_{t}-p_{r}. \label{eq20}
\end{equation}
In the present formalism, we obtain the exact solution of the EFE described in Eqs.~(\ref{eq17}), (\ref{eq18}) and (\ref{eq19}) by employing the Tolman-IV \cite{Tolman} metric ansatz defined as
\begin{equation}	
	e^{2\lambda}=\frac{1+2ar^{2}}{(1+ar^{2})(1+br^{2})}, \label{eq21}
\end{equation}
where, $a$ and $b$ both are constants with dimensions $Km^{-2}$. Now, substituting Eq.~(\ref{eq21}) in Eq.~(\ref{eq17}), we obtain the analytical expression of the energy density of the form:
\begin{equation}
	\rho=\frac{a(3-7br^{2})+a^{2}(2r^{2}-6br^{4})-3b}{8\pi(1+2ar^{2})^{2}}. \label{eq22}
\end{equation}
In the present scenario, we have considered the MIT bag model EoS in presence of non-zero strange quark mass $(m_{s}\neq0)$ and $n$-dependent bag constant as shown in Eq.(\ref{eq11}). Using Eq.~(\ref{eq22}) in Eq.(\ref{eq11}), we obtain the radial component of pressure $(p_{r})$ in the following form:
\begin{equation}
	p_{r}=\frac{1}{3}\Bigg[\frac{a(3-7br^{2})+a^{2}(2r^{2}-6br^{4})-3b}{8\pi(1+2ar^{2})^{2}}-4B'\Bigg].\label{eq23}
\end{equation} 
Substituting Eq.~(\ref{eq23}) in Eq.~(\ref{eq18}), we obtain the $g_{tt}$ metric potential $(\nu)$ as
\begin{eqnarray}
	\nu=\frac{1}{6b(a-b)}\Bigg[b(a+b+16\pi B')\log{(1+ar^{2})}-\nonumber\\b(b-a)\log{(1+2ar^{2})}-\nonumber\\(5ab-3b^{2}+32\pi a B'-16\pi b B')\log{(1+br^{2})}\Bigg].\label{eq24}
\end{eqnarray} 
Now, substituting Eq.~(\ref{eq24}) in Eq.~(\ref{eq19}), we obtain
\begin{eqnarray}
	p_{t}=\frac{1}{72\pi(1+ar^{2})(1+br^{2})(1+2ar^{2})^{3}}\Bigg[9b^{2}r^{2}\nonumber\\+32\pi B'(8\pi B'r^{2}-3)+b(48\pi B'r^{2}-9)\nonumber\\+A+B+C+D\Bigg], \label{eq25}
\end{eqnarray}  
where, $A,~B$ and $C$ are constants and, \\
$A=4a^{4}r^{6}(4-27br^{2}-320\pi B'r^{2}+9b^{2}r^{4}+96\pi B'br^{4}+1024\pi^{2}B'^2r^{4})$ \\
$B=a^{3}(30r^{4}-64r^{6}(4b+47\pi B')+r^{8}(90b^{2}+640\pi b B'+8192\pi^{2}B'^{2}))$ \\
$C=a(9-816\pi B'r^{2}+57b^{2}r^{4}+2048\pi^{2}B'^{2}r^{4}+4br^{2}(64\pi B'r^{2}-15))$ \\
$D=a^{2}r^{2}(21-2432\pi B'r^{2}+106b^{2}r^{4}+6144\pi^{2}B'^{2}r^{4}+br^{2}(544\pi B'r^{2}-201))$ \\
The total active gravitational mass enclosed within the sphere of radius $R$ is expressed as:
\begin{equation}
	m(r)=4\pi \int_{0}^{R}\rho r^{2} dr. \label{eq26}
\end{equation}
\section{Boundary condition} \label{sec4}
The vacuum and flat exterior space-time is characterised by the Schwarzschild solution \cite{Schwarzschild} and it is expressed as
\begin{equation}
	ds^2=-\Bigg(1-\frac{2M}{r}\Bigg)dt^2+\Bigg(1-\frac{2M}{r}\Bigg)^{-1}dr^2+r^2(d\theta^2+sin^2\theta d\phi^2), \label{eq27}
\end{equation}
where, $M$ is the total mass of the stellar configuration. The continuity of the metric potentials at the stellar surface $(r=R)$ is necessary to evaluate the parameters present in the model. Following this notion, we match the interior solutions with the exterior Schwarzschild solution as:
\begin{equation}
	e^{-2\lambda(R)}=1-\frac{2M}{R}, \label{eq28}
\end{equation} 
and 
\begin{equation}
	e^{2\nu(R)}=1-\frac{2M}{R}. \label{eq29}
\end{equation} 
Again, the condition that the radial pressure $(p_{r})$ must vanish at the stellar radius $(r=R)$ is represented as
\begin{equation}
	p_{r}(r=R)=0. \label{eq30}
\end{equation}
Using Eqs.~(\ref{eq28}), (\ref{eq29}) and (\ref{eq30}), we obtain the expressions for the constants in the following forms:
\begin{equation}
	a=\frac{-2M-bR^{3}}{R^{2}(4M-R+bR^{3})}. \label{eq31}
\end{equation}
and 
\begin{eqnarray}
	b=\frac{1}{2R^{6}}\Big(-3MR^{3}-16\pi B'R^{6}+R^{3}\Big(12MR-23M^{2}+\nonumber\\96\pi B'MR^{3}-64\pi B'R^{4}+256\pi^{2}B'^{2}R^{6}\Big)\Big)^\frac{1}{2}, \label{eq32}
\end{eqnarray}

\section{Deriving mass-radius relationship from TOV equation} \label{sec5} To obtain the maximum allowed range of mass and the associated radius, we have numerically solved the TOV equations \cite{Tolman,Oppenheimer} using MIT EoS with $B(n)=57.55,~70$ and $95.11~MeV/fm^{3}$ corresponding to $n=0.66,~0.578$ and $0.464~fm^{-3}$ respectively, along with different choices of strange quark mass, $m_{s}=0,~50$ and $100~MeV$.   
\begin{figure}[h]
	\centering
	\includegraphics[width=8cm]{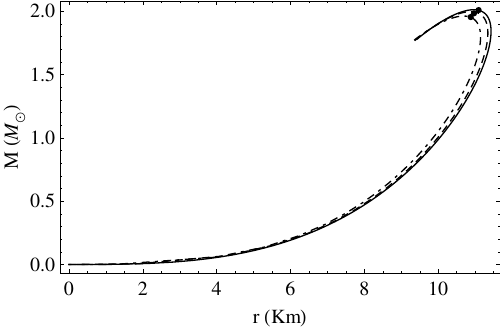}
	\caption{Mass-radius plot for $B(n)=57.55~MeV/fm^{3}$. Here, solid, dashed and dot-dashed lines correspond to $m_{s}=0,~50$ and $100~MeV$ respectively. The maximum mass points are indicated with black solid circles.}
	\label{fig3a}
\end{figure} 
\begin{figure}[h]
	\centering
	\includegraphics[width=8cm]{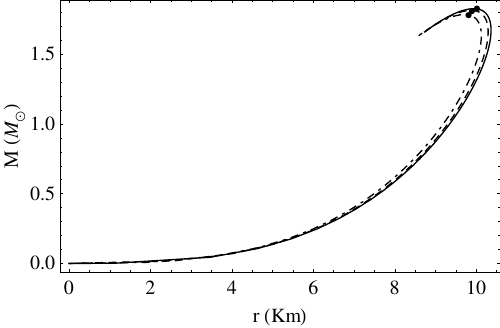}
	\caption{Mass-radius plot for $B(n)=70~MeV/fm^{3}$. Here, solid, dashed and dot-dashed lines correspond to $m_{s}=0,~50$ and $100~MeV$ respectively. The maximum mass points are indicated with black solid circles.}
	\label{fig3}
\end{figure} 
\begin{figure}[h]
	\centering
	\includegraphics[width=8cm]{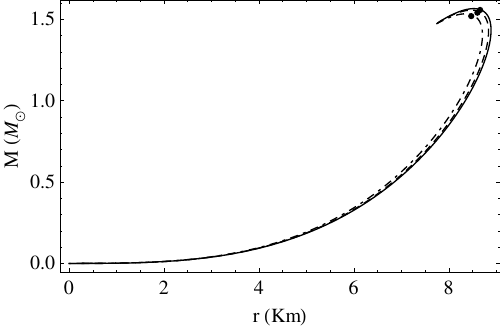}
	\caption{Mass-radius plot for $B(n)=95.11~MeV/fm^{3}$. Here, solid, dashed and dot-dashed lines correspond to $m_{s}=0,~50$ and $100~MeV$ respectively. The maximum mass points are indicated with black solid circles.}
	\label{fig3aa}
\end{figure}
From Figs.~(\ref{fig3a}), (\ref{fig3}) and (\ref{fig3aa}), it is evident that as $m_{s}$ increases the maximum mass and the radius both decrease when $B(n)$ or $n$ is fixed. This may be attributed to the fact that as the mass of the strange quark increases, the associated Fermi energy increases as the energy of the degenerate fermions depends on the mass and Fermi momentum of the strange quark. This increased energy leads to a softer EoS. Moreover, An increase in the strange quark mass reduces the quark number density for a given pressure, as more massive quarks take up more energy. This means that for a fixed pressure, fewer quarks can be packed into a given volume. As a result, the energy density increases rapidly for higher baryon number densities. This increased energy density leads to a quicker gravitational collapse which ultimately describes a smaller value of the maximum mass and the associated radius. Therefore, with increasing $m_{s}$ the EoS becomes softer and the maximum mass decreases which is evident from Figs.~(\ref{fig3a}), (\ref{fig3}) and (\ref{fig3aa}). We have tabulated the value of maximum mass and corresponding radius in Tables~\ref{tab3a}, \ref{tab3} and \ref{tab3aa} for $B(n)=57.55,~70$ and $95.11~MeV/fm^{3}$ and $m_{s}=0,~50$ and $100~MeV$. On the other hand, from Tables~\ref{tab3a}, \ref{tab3} and \ref{tab3aa}, it is evident that when $m_{s}$ is fixed, maximum mass and radius both increase with increase of $n$ as $B(n)$ decreases with increasing $n$. 
\begin{table}[h!]
	\centering
	\caption{Evaluation of maximum mass and radius for bag constant $B(n)=57.55~MeV/fm^{3}$ and different choices of strange quark mass $(m_{s})$.}
	\label{tab3a}
	\begin{tabular}{ccc}
		\hline
		Strage quark mass  & Maximum mass & Radius \\
		$(m_{s})~(MeV)$	& $(M_{max})~(M_{\odot})$ & (Km) \\
		\hline
		0 & 2.01 & 10.96 \\
		50 & 1.99 & 10.88 \\
		100 & 1.96 & 10.69 \\
		\hline
	\end{tabular}
\end{table} 
\begin{table}[h!]
	\centering
	\caption{Evaluation of maximum mass and radius for bag constant $B(n)=70~MeV/fm^{3}$ and different choices of strange quark mass $(m_{s})$.}
	\label{tab3}
	\begin{tabular}{ccc}
		\hline
		Strage quark mass  & Maximum mass & Radius \\
		$(m_{s})~(MeV)$	& $(M_{max})~(M_{\odot})$ & (Km) \\
		\hline
		0 & 1.82 & 9.94 \\
		50 & 1.81 & 9.87 \\
		100 & 1.78 & 9.72 \\
		\hline
	\end{tabular}
\end{table}  
\begin{table}[h!]
	\centering
	\caption{Evaluation of maximum mass and radius for bag constant $B(n)=95.11~MeV/fm^{3}$ and different choices of strange quark mass $(m_{s})$.}
	\label{tab3aa}
	\begin{tabular}{ccc}
		\hline
		Strage quark mass  & Maximum mass & Radius \\
		$(m_{s})~(MeV)$	& $(M_{max})~(M_{\odot})$ & (Km) \\
		\hline
		0 & 1.56 & 8.53 \\
		50 & 1.55 & 8.48 \\
		100 & 1.53 & 8.35 \\
		\hline
	\end{tabular}
\end{table}  
Here, we have predicted the radii of a wide class of compact stars for different choices of $m_{s}$ and they are tabulated in Table~\ref{tab4}. Moreover, using the predicted radii, we have tabulated the central density $(\rho_{0})$, surface density $(\rho_{s})$ and central pressure $(p_{0})$ for the known compact stars and they are tabulated in Table~\ref{tab5}.
\begin{table}[h!]
	\centering
	\caption{Tabulation of radii prediction of known compact stars for $B(n)=70~MeV/fm^{3}$ and different choices of strange quark mass $(m_{s})$.}
	\label{tab4}
	\begin{tabular}{ccccc}
		\hline
		Compact Stars & Measure mass $(M_{\odot})$ & Measured radius $(Km)$ & Strange quark mass $(m_{s})$ & Predicted radius $(Km)$ \\
		\hline
		4U 1820-30 \cite{Guver} & $1.58^{+0.06}_{-0.06}$ & $9.1^{+0.4}_{-0.4}$ & 150 & 9.94 \\
		4U 1608-52 \cite{Guver1} & $1.74^{+0.14}_{-0.14}$ & $9.3^{+1.0}_{-1.0}$ & 100 & 9.20 \\
		EXO 1745-248 \cite{Ozel} & 1.7 & 9.0 & 100 & 8.94 \\
		Cen X-3 \cite{Rawls} & $1.49^{+0.08}_{-0.08}$ & $9.178^{+0.13}_{-0.13}$ & 150 & 9.89 \\
		Her X-1 \cite{Abubekerov} & $0.85^{+0.15}_{-0.15}$ & $8.1^{+0.41}_{-0.41}$ & 50 & 8.91 \\
		\hline
	\end{tabular}
\end{table}
\begin{table}[h!]
	\centering
	\caption{Tabulation of physical parameters of the known compact stars.}
	\label{tab5}
	\begin{tabular}{cccc}
		\hline
		Compact object & Central density $(\rho_{0})$ & Surface density $(\rho_{s})$ & Central pressure $(p_{0})$ \\ 
		& $(gm/cc)$ & $(gm/cc)$ & $(dyn/cm^{2})$ \\
		\hline
		4U 1820-30 & $1.59\times10^{15}$ & $0.54\times10^{15}$ & $3.15\times10^{35}$\\
		4U 1608-52 & $1.59\times10^{15}$ & $0.54\times10^{15}$ & $3.15\times10^{35}$\\
		EXO 1745-248 & $1.59\times10^{15}$ & $0.54\times10^{15}$ & $3.15\times10^{35}$ \\
		Cen X-3 & $1.59\times10^{15}$ & $0.54\times10^{15}$ & $3.15\times10^{35}$\\
		Her X-1 & $1.59\times10^{15}$ & $0.54\times10^{15}$ & $3.15\times10^{35}$\\
		\hline
	\end{tabular}
\end{table}
\newpage
\section{Physical features of the present model}\label{sec6} To emerge as a physically realistic stellar model, the present theoretical structure must adhere to certain regularity and viable conditions of different stellar parameters {\it viz.}, energy density $(\rho)$, radial $(p_{r})$ and tangential $(p_{t})$ pressures, pressure anisotropy $(\Delta)$, causality and energy conditions. In reference to the above, we consider the compact star candidate Low-Mass X-ray Binary (LMXB) 4U 1820-30 having a mass of $1.58~M_{\odot}$ and radius of $9.1$ Km \cite{Guver} to investigate the physical acceptability of the proposed model. In the following segment, we have graphically represented the radial variations of energy density $(\rho)$, radial $(p_{r})$ and tangential $(p_{t})$ pressures, pressure anisotropy $(\Delta)$ for different choices of strange quark mass $(m_{s})$ to study the impact of finite value of $m_{s}$ on the characteristic parameters. For such analysis, we have considered $n=0.578~fm^{-3}$ and $B(n)=70~MeV/fm^{3}$.
\begin{figure}[h]
	\centering
	\includegraphics[width=8cm]{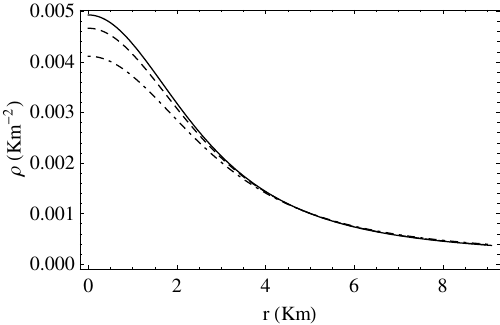}
	\caption{Radial variation of energy density $(\rho)$. Here, solid, dashed and dot-dashed lines correspond to $m_{s}=0,~50$ and $100~MeV$ respectively.}
	\label{fig4}
\end{figure}
\begin{figure}[h]
	\centering
	\includegraphics[width=8cm]{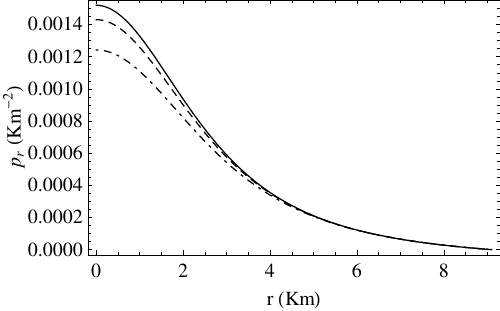}
	\caption{Radial variation of radial component of pressure $(p_{r})$. Here, solid, dashed and dot-dashed lines correspond to $m_{s}=0,~50$ and $100~MeV$ respectively.}
	\label{fig5}
\end{figure}
\begin{figure}[h]
	\centering
	\includegraphics[width=8cm]{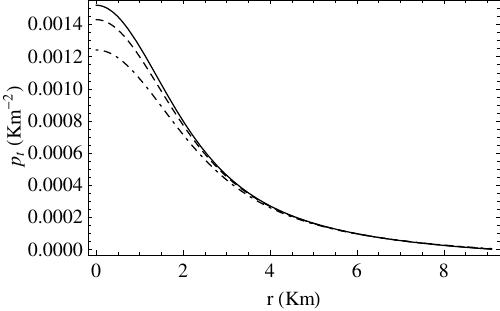}
	\caption{Radial variation of tangential component of pressure $(p_{t})$. Here, solid, dashed and dot-dashed lines correspond to $m_{s}=0,~50$ and $100~MeV$ respectively.}
	\label{fig6}
\end{figure}
\begin{figure}[h]
	\centering
	\includegraphics[width=8cm]{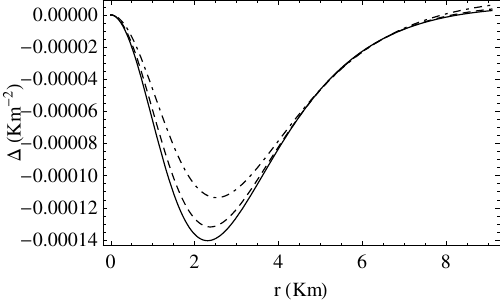}
	\caption{Radial variation of pressure anisotropy $(\Delta)$. Here, solid, dashed and dot-dashed lines correspond to $m_{s}=0,~50$ and $100~MeV$ respectively.}
	\label{fig7}
\end{figure}
\\
From Figs.~(\ref{fig4}), (\ref{fig5}) and (\ref{fig6}), we note that the radial variations of the energy density and pressure profiles display a monotonically decreasing nature from centre to the stellar surface which points toward a physically acceptable stellar model. Moreover, from Fig.~(\ref{fig7}), we note that the pressure anisotropy in the present formalism is negative implying an attractive nature. The decrease in energy density and pressure profiles with increasing strange quark mass can be attributed to the added inertia introduced by the heavier quarks, which makes the system less compact. Heavier quarks require more energy to be confined, reducing the overall energy density and pressure in the star. As the strange quark mass increases, the pressure anisotropy rises due to the fact that the radial and tangential pressure components respond differently to the change in quark mass. This anisotropy is an intriguing feature of the model because it reflects how the system's internal forces become increasingly imbalanced, potentially due to the competing effects of gravitational forces and quark interactions. The presence of heavier quarks may enhance the anisotropy by influencing the spatial distribution of forces within the star, making the structure more anisotropic.
\subsection{Causality condition} 
\begin{figure}[h]
	\centering
	\includegraphics[width=8cm]{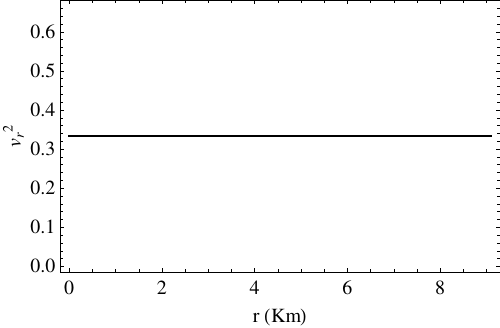}
	\caption{Radial variation of radial sound velocity $(v_{r}^{2})$. Here, solid, dashed and dot-dashed lines correspond to $m_{s}=0,~50$ and $100~MeV$ respectively.}
	\label{fig8}
\end{figure}

\begin{figure}[h]
	\centering
	\includegraphics[width=8cm]{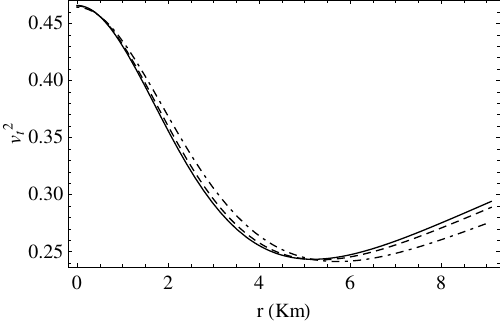}
	\caption{Radial variation of tangential sound velocity $(v_{t}^{2})$. Here, solid, dashed and dot-dashed lines correspond to $m_{s}=0,~50$ and $100~MeV$ respectively.}
	\label{fig9}
\end{figure}

To develop a realistic model for an anisotropic compact star, one key method for characterising the dense matter in its interior is through analysing the propagation velocities of sound waves. The radial and tangential sound velocities are given by $v_{r}^{2}=\frac{dp_{r}}{d\rho}$ and $v_{t}^{2}=\frac{dp_{t}}{d\rho}$, respectively, where $\rho$, $p_{r}$, and $p_{t}$ have been previously defined. In this analysis, we employ natural units where $\hbar=c=1$. The causality condition requires that both sound velocities remain within the upper limit of $v_{r}^{2}\leq1$ and $v_{t}^{2}\leq1$. Additionally, thermodynamic stability mandates that $v_{r}^{2}>0$ and $v_{t}^{2}>0$. Consequently, both constraints must be satisfied, implying that $0<v_{r}^{2}\leq1$ and $0<v_{t}^{2}\leq1$ throughout the stellar structure. To avoid complex calculations, the radial dependence of these sound velocities is shown graphically in Figs.~(\ref{fig8}) and (\ref{fig9}). 
\\
Interestingly, since the EoS is represented by Eq.~(\ref{eq11}), the radial sound velocity $v_{r}^{2}=\frac{dp_{r}}{d\rho}$ is always $\frac{1}{3}$ irrespective of the choice of $m_{s}$. However, the influence of non-zero strange quark mass is eminent in the case of $v_{t}^{2}$ as shown in Fig.~(\ref{fig9}). We note that, the causality condition is well met in the present model. 
\subsection{Energy conditions} In gravitational theory, energy conditions are constraints applied to matter distributions to ensure a physically consistent energy-momentum tensor. These conditions are crucial for exploring the properties of matter distributions without requiring detailed knowledge of their internal composition. This makes it possible to infer the physical characteristics of extreme phenomena, such as the formation of geometric singularities or the gravitational collapse following a merger, without direct information on pressure or energy density. Essentially, evaluating energy conditions is an algebraic problem, as described in \cite{Kolassis}, and can be understood as an eigenvalue problem related to the energy-momentum tensor.
\begin{figure}[h!]
	\centering
	\includegraphics[width=8cm]{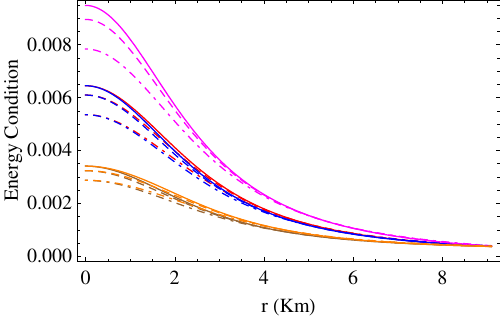}
	\caption{Radial variation of Energy conditions. Here, red, blue, magenta, brown and orange lines represent $(\rho+p_{r})$, $(\rho+p_{t})$, $(\rho+p_{r}+2p_{t})$, $(\rho-p_{r})$ and $(\rho-p_{t})$ respectively. Additionally, the solid, dashed and dot-dashed lines correspond to $m_{s}=0,~50$ and $100~MeV$ respectively.}
	\label{fig10}
\end{figure}
In a four-dimensional spacetime, analysing energy conditions leads to solving the roots of a polynomial of degree four, which becomes complex due to the presence of analytical solutions for the eigenvalues. Despite the challenges in deriving a general solution, a physically realistic matter distribution must simultaneously satisfy the dominant energy condition, strong energy condition, weak energy condition, and null energy condition--collectively known as the energy conditions \cite{Kolassis, Hawking, Wald}. In this work, we have assessed the energy conditions \cite{Brassel, Brassel1} for the present stellar model as expressed in the form below:
\begin{enumerate}
	\item Null energy condition (NEC): $\rho+p_{r}\geq0, \rho+p_{t}\geq0 $. \\
	\item Weak energy condition (WEC): $\rho\geq0, \rho+p_{r}\geq0, \rho+p_{t}\geq0 $. \\
	\item Strong energy condition (SEC): $\rho+p_{r}\geq0, \rho+p_{t}\geq0, \rho+p_{r}+2p_{t}\geq0$. \\
	\item Dominant Energy Condition (DEC): $\rho\geq0, \rho-p_{r}\geq0, \rho-p_{t}\geq0 $.
\end{enumerate}
Fig.~(\ref{fig10}) illustrates that the present model satisfies all the necessary energy conditions throughout the stellar interior in presence of finite value of strange quark mass $(m_{s})$ and baryon number density dependent $B$.
\section{Stability analysis}\label{sec7} In this scenario, we have employed the following methods to assess the stability of the present model: \\
(i) Generalised form of TOV equation, \\
(ii) Cracking condition based on Herrera's proposal, \\
(iii) Lagrangian perturbation and \\
(iv) Radial variation of the adiabatic index. 
\subsection{Generalised form of TOV equation} 
\begin{figure}[h]
	\centering
	\includegraphics[width=8cm]{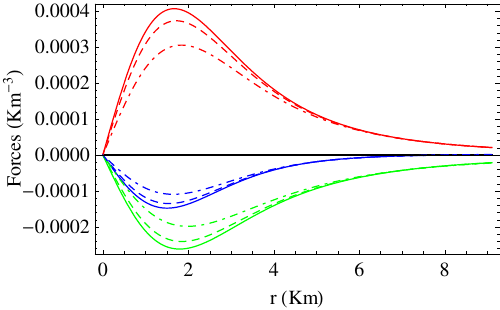}
	\caption{Radial variation of different forces. Here, red, blue, and green lines represent $F_{h}$, $F_{a}$ and $F_{g}$ respectively. Here, $n=0.578~fm^{-3}$ and the solid, dashed and dot-dashed lines correspond to $m_{s}=0,~50$ and $100~MeV$ respectively.}
	\label{fig11}
\end{figure}

To analyse the stability of an anisotropic compact object, it is crucial to consider the effects of various forces acting on it. Specifically, stability is assessed through three force components: (i) the gravitational force ($F_{g}$), (ii) the hydrostatic force ($F_{h}$), and (iii) the anisotropic force ($F_{a}$). The system reaches equilibrium when these forces collectively balance each other. In this framework, we have applied the generalised Tolman-Oppenheimer-Volkoff (TOV) equation \cite{Tolman,Oppenheimer} expressed as follows:
\begin{equation}
	-\frac{M_{G}(r)(\rho+p_{r})}{r^2}e^{\lambda-\nu}-\frac{dp_{r}}{dr}+\frac{2\Delta}{r}=0, \label{eq33}
\end{equation} 
where, $M_{G}$ is referred to as the active gravitational mass derived from the mass formula of Tolman-Whittaker \cite{Gron} given as: 
\begin{equation}
	M_{G}(r)=r^{2}\nu'e^{\nu-\lambda}. \label{eq34}
\end{equation}
Substituting Eq.~(\ref{eq34}) in Eq.~(\ref{eq33}), we obtain
\begin{equation}
	-\nu'(\rho+p_{r})-\frac{dp_{r}}{dr}+\frac{2\Delta}{r}=0. \label{eq35}
\end{equation}  
Here, 
\begin{equation}
	F_{g}=-\nu'(\rho+p_{r}),  \label{eq36} 
\end{equation}
\begin{equation}
	F_{h}=-\frac{dp_{r}}{dr},  \label{eq37}
\end{equation}	
and
\begin{equation}
	F_{a}=\frac{2\Delta}{r}.  \label{eq38}
\end{equation}
Therefore, Eq.~(\ref{eq35}) can be written as
\begin{equation}
	F_{g}+F_{h}+F_{a}=0. \label{eq39}
\end{equation}
Using the expressions from Eqs.~(\ref{eq22}), (\ref{eq23}), (\ref{eq24}) and (\ref{eq25}), we have computed the necessary force components. Additionally, to avoid the complex mathematical analysis, we have opted for the graphical representation of Eq.~(\ref{eq39}) which is illustrated in Fig.~(\ref{fig11}). 
From Fig.~(\ref{fig11}), it is evident that the sum of all the force components nullifies within the stellar structure implying the static equilibrium of the present model in presence of nonzero $m_{s}$ and  baryon number dependent $B$. 
\subsection{Cracking condition based on Herrera's proposal} Anisotropic stellar models should maintain stability when subject to fluctuations in their physical parameters. Herrera \cite{Herrera1} introduced the concept of the "cracking" to assess the stability of such models. Building on this idea, Abreu et al. \cite{Abreu} proposed a stability criterion for anisotropic stellar models. According to their approach, the model remains stable if the squares of the radial ($v_{r}^{2}$) and the tangential velocities ($v_{t}^{2}$) satisfy the following condition
\begin{equation}
	0\leq|v_{t}^{2}-v_{r}^{2}|\leq 1. \label{eq40}
\end{equation}
\begin{figure}[h]
	\centering
	\includegraphics[width=8cm]{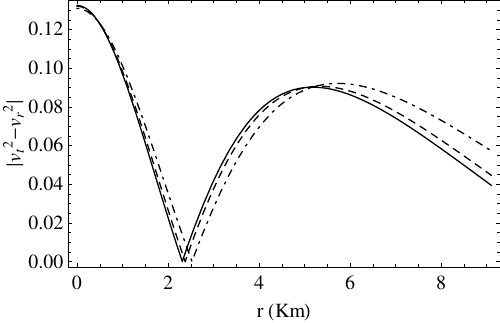}
	\caption{Radial variation of $|v_{t}^{2}-v_{r}^{2}|$. Here, $n=0.578~fm^{-3}$ and the solid, dashed and dot-dashed lines correspond to $m_{s}=0,~50$ and $100~MeV$ respectively.}
	\label{fig12}
\end{figure}
Fig.~(\ref{fig12}) depicts that the stability condition put forward by Abreu et al. \cite{Abreu} is well maintained within the stellar configuration in presence of finite $m_{s}$ and $B(n)$. 
\subsection{Lagrangian perturbation} To investigate the stability of our model under small radial oscillations, we present a graphical analysis of the variation in the Lagrangian change of radial pressure at the surface of a compact object as a function of the frequency $(\omega^{2})$. Pretel \cite{Pretel} introduced a method for illustrating the frequency dependence of the Lagrangian perturbation. The coupled equations governing the radial oscillations are formulated as follows:
\begin{eqnarray}
	\frac{d\eta}{dr}=-\frac{1}{r}(3\eta+\frac{\Delta p_{r}}{\Gamma p_{r}})+\frac{d\nu}{dr}\eta,\label{eq41} \\ 
	\frac{d\Delta p_{r}}{dr}=\eta\Big(\frac{\omega^2}{c^2}e^{2(\lambda-\nu)}(\rho+p_{r})r-4\frac{dp_{r}}{dr}\nonumber \\-\frac{8\pi G}{c^4}(\rho+p_{r})e^{2\lambda}rp_{r}+r(\rho+p_{r})(\frac{d\nu}{dr})\Big)\nonumber\\-\Delta p_{r}\Big(\frac{d\nu}{dr}+\frac{4\pi G}{c^4}(\rho+p_{r})re^{2\lambda}\Big), \label{eq42}
\end{eqnarray} 
Here, $\eta$ represents the eigenfunction corresponding to the radial component of the Lagrangian displacement, defined as $\eta = \frac{\delta(r)}{r}$. In this formulation, $\eta$ is normalized such that $\eta(0) = 1$. To eliminate the central singularity in Eq.~(\ref{eq41}), the term involving $(\frac{1}{r})$ must vanish as $r \rightarrow 0$. Following this proposition, we arrive at the following criterion:
\begin{equation}
	\Delta p_{r}=-3\Gamma\eta p_{r}=3\Bigg(\frac{\rho+p_{r}}{p_{r}}\frac{dp_{r}}{d\rho}\Bigg)\eta p_{r}. \label{eq43}
\end{equation}
Moreover, the change in radial component of pressure $\Delta p_{r}$, due to the introduction of frequency dependence of the Lagrangian perturbation, must also vanish at the stellar surface, i.e., 
\begin{equation}
	\Delta p_{r}|_{r=R}=0. \label{eq44}
\end{equation} 
\begin{figure}[h]
	\centering
	\includegraphics[width=8cm]{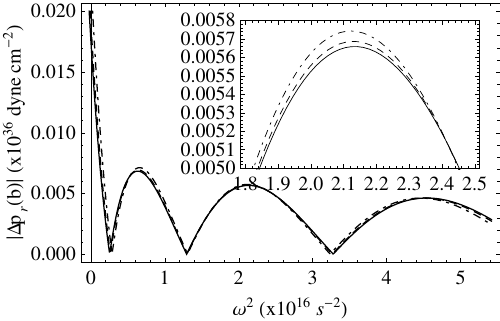}
	\caption{Variation of Lagrangian perturbation with respect to normal mode frequencies. Here, $n=0.578~fm^{-3}$ and the solid, dashed and dot-dashed lines correspond to $m_{s}=0,~50$ and $100~MeV$ respectively.}
	\label{fig13}
\end{figure}
As shown in Fig.~(\ref{fig13}), $\omega^{2}$ remains positive for all normal modes of radial oscillation in this model, and the correct normal frequency modes are indicated by the minima of the plot. Based on this criterion, we can conclude that the model is stable against small radial oscillatory perturbations also. 
\subsection{Radial variation of the adiabatic index} For a relativistic stellar structure with anisotropic perfect fluid distribution, the expression of adiabatic index is written as
\begin{equation}
	\Gamma=\frac{\rho+p_{r}}{p_{r}}\frac{dp_{r}}{d\rho}=\frac{\rho+p_{r}}{p_{r}}v_{r}^{2}. \label{eq45}
\end{equation}
Heintzmann and Hillebrandt \cite{Heintzmann} established that the stability condition for an isotropic stellar model is given by $\Gamma > \frac{4}{3}$ and termed the Newtonian limit. For an anisotropic star, where both radial pressure ($p_{r}$) and tangential pressure ($p_{t}$) are present, Chan et al. \cite{Chan} modified this stability condition, which is expressed as follows:
\begin{equation}
	\Gamma>\Gamma'_{max}, \label{eq46}
\end{equation}
where, 
\begin{equation}
	\Gamma'_{max}=\frac{4}{3}-\Bigg[\frac{4}{3}\frac{(p_{r}-p_{t})}{|p'_{r}|r}\Bigg]_{max}. \label{eq47}
\end{equation}
\begin{figure}[h]
	\centering
	\includegraphics[width=8cm]{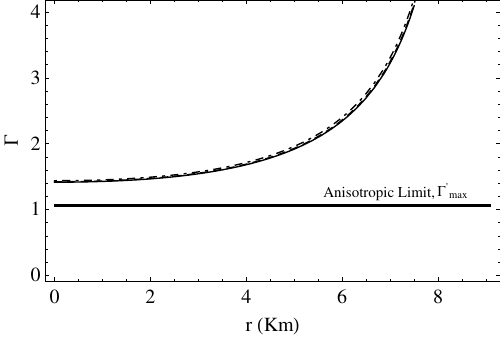}
	\caption{Radial variation of adiabatic index. Here, $n=0.578~fm^{-3}$ and the solid, dashed and dot-dashed lines correspond to $m_{s}=0,~50$ and $100~MeV$ respectively.}
	\label{fig14}
\end{figure}
Interestingly, in the present scenario we recall that the pressure anisotropy shows an attractive nature. Hence, the term in the bracket in Eq.~(\ref{eq47}) is positive in nature and the whole numerical value of $\Gamma'_{max}$ is less that $\frac{4}{3}$. However, the adiabatic index in the present model adheres to the condition given in Eq.~(\ref{eq46}) which demonstrates a stable and physically acceptable stellar model. 
\section{Conclusion}\label{sec8}
In this paper, we have studied the impact of strange quark mass $(m_{s})$ and baryon number density $(n)$ on the structure and stability of strange stars. To obtain a theoretical relativistic stellar configuration, we have considered the Tolman-IV metric ansatz \cite{Tolman} and obtained the exact set of solutions of the EFE. The modeling of compact stars is significantly influenced by the choice of the EoS, especially in the high-density regime. In the context of SS modeling, the MIT bag model is commonly employed to describe the deconfined quark phase at such extreme densities. However, following CERN experimental findings, it is well-recognised that in the high-density regime $(\sim 10^{15} gm/cc)$, a phase transition from hadronic matter to a Quark-Gluon Plasma (QGP) may occur. The standard MIT bag model, which assumes a constant bag constant $(B)$, may not be sufficient to capture these phase transition characteristics. Density dependent $B$ or baryon number dependent $B$ may explain such phase transitions, more correctly. To develop a more realistic EoS, we introduce the presence of strange quark mass $(m_{s})$ and Wood-Saxon-type parametrisation \cite{Wood} for $B$, where it is dependent on the baryon number density $(n)$ \cite{Burgio}. The composite effects of $m_{s}$ and $n$ significantly modify $B$ as expressed in Eq.~(\ref{eq12a}). Notably, considering the zero external pressure, Madsen \cite{Madsen} showed that for a stable quark matter composed of three flavour quarks, the bag constant lies within the range $57.55\leq B\leq95.11~MeV/fm^{3}$. Now, using Eq.~(\ref{eq12a}), we have determined the baryon number densities for this range and are tabulated in Table~\ref{tab1}. From the perspective of thermodynamic viability, we have calculated the energy per baryon number $\mathcal{E_{B}}$ and plotted the variation of $\mathcal{E_{B}}$ with respect to $m_{s}$ for different choices of bag constant $(B)$ in Fig.~(\ref{fig1}). From Fig.~(\ref{fig1}), it is clear that the strange quark mass determines whether strange quark matter (SQM) is stable, metastable, or unstable. Table~\ref{tab2} presents the allowed ranges for $m_{s}$ based on the different values of $B$ used. Additionally, we have noted some interesting observations regarding the variation of energy per baryon $(\mathcal{E_B})$ as a function of baryon number density $(n)$, illustrated in Fig.(\ref{fig2}). This figure highlights how $m_{s}$ influences the $(\mathcal{E_B})$ vs. $n$ relationship. Notably, for $m_{s}=0,~50$ and $100~MeV$, the matter exhibits unstable behaviour with respect to the quark phase, when $n<0.464,~0.473$ and $0.50~fm^{-3}$ respectively. It signifies that matter remains stable in the hadronic phase for such parametric choice. A potential phase transition from hadronic to quark matter is initiated when $n\geq0.464,~0.473$ and $0.50~fm^{-3}$ for the respective $m_{s}$ values. It is possible that this phase coexistence between hadrons and quarks results in a metastable state. Metastability is observed in the ranges $0.464\leq n\leq0.48~fm^{-3}$, $0.473\leq n\leq0.49~fm^{-3}$ and $0.50\leq n\leq0.51~fm^{-3}$ for $m_{s}=0,~50$ and $100~MeV$, respectively. It is interesting to note that stability window decreases in terms of baryon number density $n$, when $m_{s}$ increases. On the other hand, complete conversion from hadronic to quark phase occurs at higher values of $n$ when $m_{s}$ increases. This may be attributed to the fact that at higher baryon densities $(n)$, the system possesses greater energy. However, when $m_{s}$ is increased, this higher density provides sufficient energy to overcome the energy barrier for the production of strange quarks, thereby initiating the phase transition to quark matter. As the baryon number density increases beyond $n\geq0.482,~0.489$ and $0.51~fm^{-3}$, the energy per baryon reaches a plateau, signalling that the hadronic matter has fully converted into quark matter, and the system achieves a stable state. This behaviour is physically reasonable as at lower baryon number densities the phase space states remain largely unoccupied, allowing quarks to fill available states below the Fermi level. However, as the baryon number density $n$ increases beyond a certain threshold value, these free states become fully occupied by quarks, leading to a degenerate system. At this point, no additional quarks can be formed, and the system reaches an equilibrium configuration. As a result, the energy per baryon stabilises and no longer varies with increasing $n$. In this state, the star becomes incompressible and cannot undergo further compression. Furthermore, Table~\ref{tab1} shows that selecting $B(n)=70~MeV/fm^{3}$ yields $n=0.578~fm^{3}$, which is consistent with the expected range of $n$ for different $m_{s}$ values. Using the bag value and particular choices of $m_{s}$, we have solved the TOV equations \cite{Tolman,Oppenheimer} to determine the maximum mass and the associated radius in this model. From Fig.~(\ref{fig3}), it is evident that the well-acknowledged result that the EoS becomes softer upon increasing $m_{s}$ is well maintained in our model. From Table~\ref{tab3}, it is observed that the maximum mass and corresponding radius for the current model with massless strange quarks are 1.82 $M_{\odot}$ and 9.94 Km, respectively. However, as the strange quark mass increases, the maximum mass and radius decreases. Furthermore, we have  predicted the radii for various compact stars based on different $m_{s}$ values, as shown in Table~\ref{tab4}. Using these radii, we have also tabulated the central density $(\rho_{0})$, surface density $(\rho_{s})$ and central pressure $(p_{0})$ for known compact stars in Table~\ref{tab5}. Now, to test the physical acceptability of the present model, we have considered the low-mass X-ray binary (LMXB) 4U 1820-30 having a mass of $1.58~M_{\odot}$ and radius of $9.1$ Km \cite{Guver} and utilising the Eqs.~(\ref{eq22}), (\ref{eq23}), (\ref{eq24}) and (\ref{eq25}), we have studied the fundamental characteristic properties $\it viz.,$ energy density $(\rho)$, radial $(p_{r})$ and tangential $(p_{t})$ pressure components and pressure anisotropy $(\Delta)$ for different values of strange quark mass $(m_{s})$ through the graphical representations via Figs.~(\ref{fig4}), (\ref{fig5}), (\ref{fig6}) and (\ref{fig7}). It must be noted that the monotonically decreasing energy density and pressure profiles signify a viable stellar configuration. However, the pressure anisotropy is negative in the present model which implies an attractive anisotropic nature. The reduction in energy density and pressure profiles as the strange quark mass increases can be explained by the added inertia from the heavier quarks, which leads to a less compact system. Since heavier quarks require more energy for confinement, this results in lower overall energy density and pressure within the star. Furthermore, as the strange quark mass increases, pressure anisotropy becomes more pronounced because the radial and tangential pressure components are affected differently by the increased quark mass. This growing anisotropy is an interesting aspect of the model, as it indicates an imbalance in the system's internal forces, likely due to the competing influences of gravitational forces and quark interactions. The increased mass of the quarks may amplify this anisotropy by altering the distribution of internal forces, causing the star's structure to become more anisotropic. Moreover, the causality and energy conditions are illustrated in Figs.~(\ref{fig8}), (\ref{fig9}) and (\ref{fig10}) respectively, and we note that these physical viability conditions are well satisfied in this model. The stability of the present model is assessed on the basis of generalised solution of TOV equations, the concept of cracking proposed by Herrera, Lagragian perturbation and the radial variation of adiabatic index. We have demonstrated the stability criteria through the Figs.~(\ref{fig11}), (\ref{fig12}), (\ref{fig13}) and (\ref{fig14}) respectively, and notably, all the required stability conditions are well met in this model which points toward a stable and physically acceptable stellar model. In view of the above, we may conclude that our model describes a physically realistic anisotropic stellar configuration with the interior matter composed of three flavour quarks. Moreover, the composite presence of finite strange quark $(m_{s}\neq0)$ and baryon number density $(n)$ dependent bag model simulates a more realistic EoS to investigate the structure and stability of SS.     
\section{Acknowledgements}
PKC gratefully acknowledges the support from the Inter-University Centre for Astronomy and Astrophysics (IUCAA), Pune, India under the Visiting Associateship Programme. DB is thankful to the Department of Science and Technology (DST), Govt. of India for providing the fellowship vide no: DST/INSPIRE Fellowship/2021/IF210761.


\begin{thebibliography}{99} 
	\bibitem{Shapiro} S.L. Shapiro, S.A. Teukolsky, {\it Black Holes, White Dwarfs, and Neutron Stars} (Wiley, New York, 1983).
	\bibitem{Baade} W. Baade, F. Zwicky, Proc. Natl. Acad. Sci. U.S.A {\bf 20}, 254 (1934).
	\bibitem{Oppenheimer}J.R. Oppenheimer and G.M. Volkoff, Phys. Rev. {\bf 55}, 374 (1939).
	\bibitem{Bodmer} A.R. Bodmer, Phys. Rev. D 4, 1601 (1971).
	\bibitem{Witten}  E. Witten, Phys. Rev. D {\bf 30}, 272 (1984). 
	\bibitem{Itoh} N. Itoh, Prog. Theor. Phys. {\bf 44}, 291 (1970).
	\bibitem{Madsen}  J. Madsen, {\it Physics and astrophysics of strange quark matter, in Hadrons in Dense Matter and Hadrosynthesis} (Lecture Notes in Physics), vol 516, ed. by J. Cleymans, H.B. Geyer, F.G. Scholtz (Heidelberg: Springer, 1998), p. 42.
	\bibitem{Baym} G. Baym et al., Phys. Lett. B {\bf 160}, 181 (1985).
	\bibitem{Glendenning}  N.K. Glendenning, Mod. Phys. Lett. A {\bf 5}, 2197 (1990).
	\bibitem{Alcock1} C. Alcock, E. Farhi, A. Olinto, Astrophys. J. {\bf 310}, 261 (1986).
	\bibitem{Alford} M. Alford, Ann. Rev. Nucl. Part. Sci. {\bf 51}, 131 (2001).
	\bibitem{Kapusta} J. Kapusta, {\it Finite-Temperature Field Theory} (CambridgeUniv. Press, 1994, p. 163-165).
	\bibitem{Chodos1} A. Chodos et al., Phys. Rev. D {\bf 9}, 3471 (1974).
	\bibitem{Farhi} E. Farhi and R.L. Jaffe, Phys. Rev. D {\bf 30}, 2379 (1984).
	\bibitem{Brilenkov} M. Brilenkov, M. Eingorn, L. Jenkovszky et al., J. Cosmol. Astropart. Phys. {\bf 08}, 002 (2013).
	\bibitem{Paulucci} L. Paulucci, J.E. Horvath, Physics Letters B {\bf 733}, 164 (2014).
	\bibitem{Arbanil} J.D.V. Arba\~nil, M. Malheiro, J. Cosmol. Astropart. Phys. {\bf 11}, 012 (2016).
	\bibitem{Lugones} G. Lugones, J.D.V. Arba\~nil, Phys. Rev. D {\bf 95}, 064022 (2017).
	\bibitem{Chowdhury} S.R. Chowdhury, D. Deb, S. Ray, F. Rahaman et al., Int. J. Mod. Phys. D {\bf 29}, 2050001 (2020).
	\bibitem{Maharaj} S.D. Maharaj, J.M. Sunzu, S. Ray, Eur. Phys. J. Plus. {\bf 129}, 3 (2014).
	\bibitem{Abbas} G. Abbas, S. Qaisar, A. Jawad, Astrophys. Space Sci. {\bf 359}, 57 (2015).
	\bibitem{KBG} K.B. Goswami, A. Saha, P.K. Chattopadhyay and S. Karmakar, Eur. Phys. J. C {\bf 83}, 1038 (2023).
	\bibitem{KBG1} K.B. Goswami, A. Saha, P.K. Chattopadhyay, Astrophys. Space Sci. {\bf 365}, 141 (2020).
	\bibitem{KBG2} K. B. Goswami, A. Saha, P. K. Chattopadhyay, Class. Quantum Gravity {\bf 39}, 175006 (36pp) (2022).
	\bibitem{Ruderman} R. Ruderman, Astron. Astrophys. {\bf 10}, 427 (1972).
	\bibitem{Canuto} V. Canuto, Ann. Rev. Astron. Astrophys. {\bf 12}, 167 (1974).
	\bibitem{Kippenhahn} R. Kippenhahn, A. Weigert, {\it Steller Structure and Evolution} (Springer, Berlin, 1990).
	\bibitem{Broglia} R.A. Broglia, V. Zelevinsky (eds.), {\it Fifty Years of Nuclear BCS: Pairing in Finite Systems} (World Scientific Publishing Co. Pte. Ltd., Singapore, 2013).
	\bibitem{Page} D. Page, J.M. Lattimer, M. Prakash, in {\it Novel Superfluids}, vol.2, ed. by K.H. Bennemann,J.B. Ketterson (Oxford University Press, Oxford, 2014), p.505.
	\bibitem{Sawyer} R.F. Sawyer, D.J. Scalapino, Phys. Rev. D {\bf 8}, 1260 (1973).
	\bibitem{Sawyer1} R.F. Sawyer, Phys. Rev. Lett. {\bf 29}, 382 (1972).
	\bibitem{Herrera} L. Herrera, N.O. Santos, Phys. Rep. {\bf 286}, 53 (1997).
	\bibitem{Bowers} R.L. Bowers, E.P.T. Liang, Astrophys. J. 188, 657 (1974).
	\bibitem{Heintzmann} H. Heintzmann, W. Hillebrandt, Astron. Astrophys. {\bf 38}, 51 (1975).
	\bibitem{Maurya} S.K. Maurya, Y.K. Gupta, S. Ray, B. Dayanandan, Eur. Phys. J. C {\bf 75}, 225 (2015).
	\bibitem{Maurya1} S.K. Maurya, Y.K. Gupta, B. Dayanandan, M.K. Jasim, A. Al Jamel, Int. J. Mod. Phys. D {\bf 26}, 1750002 (2017).
	\bibitem{Deb} D. Deb, S.R. Chowdhury, S. Ray, F. Rahaman, B.K. Guha, Ann. Phys. (Amsterdam) {\bf 387}, 239 (2017).
	\bibitem{Kalam} M. Kalam, F.Rahaman, S.Molla, S.M. Hossein, Astrophys. Space Sci. {\bf 349}, 865 (2014).
	\bibitem{Mak} M.K. Mak, T. Harko, Proc. Roy. Soc. Lond. A {\bf 459}, 393 (2003). 
	\bibitem{Mak1}  M.K. Mak et al., Int. J. Mod. Phys. D {\bf 11}, 207 (2002.
	\bibitem{Hernandez} H. Hernandez, L. Nunez, Can. J. Phys. {\bf 82}, 29 (2004).
	\bibitem{Carter} B. Carter, D. Langlois, Nucl. Phys. B {\bf 531}, 478 (1998). 
	\bibitem{Bhattacharjee} D. Bhattacharjee, P.K. Chattopadhyay, Eur. Phys. J. C {\bf 84}, 77 (2024).
	\bibitem{Tolman} R.C. Tolman, Phys. Rev. {\bf 55}, 364 (1939).
	\bibitem{Banerjee} S. Banerjee, Pramana {\bf 91}, 27 (2018).
	\bibitem{Bhar} P. Bhar, K.N. Singh, T. Manna, Astrophys. Space Sci. {\bf 361}, 284 (2016).
	\bibitem{Das} S. Das, B.K. Parida, K. Chakraborty, S. Ray, Int. J. Mod. Phys. D {\bf 31}, 2250053 (2022).
	\bibitem{Maharaj1} S.D. Maharaj, P. Mafa Takisa, Astrophys. Space Sci. {\bf 343}, 569 (2012).
	\bibitem{Takisa} P. Mafa Takisa, S.D. Maharaj, Gen. Relativ. Gravit. {\bf 45}, 1951 (2013).
	\bibitem{Takisa1} P. Mafa Takisa, S.D. Maharaj, S. Ray, Astrophys. Space Sci. {\bf 354}, 2120 (2014).
	\bibitem{Thirukkanesh} S. Thirukkanesh, S.D. Maharaj, Class. Quantum Gravity {\bf 25}, 235001 (2008).
	\bibitem{Thirukkanesh1} S. Thirukkanesh, A. Kaisavelu, M. Govender, Eur. Phys. J. C {\bf 80}, 214 (2020).
	\bibitem{Rocha} L.S. Rocha et al., Astron. Notes {\bf 340}, 180 (2019).
	\bibitem{Reinhardt} H. Reinhardt, B. V. Dang, Phys. Lett. B {\bf 173}, 473 (1986).
	\bibitem{Burgio} G.F. Burgio et al., Phys. Lett. B {\bf 526}, 19 (2002).
	\bibitem{Wood} R. D. Woods and D. S. Saxon, Phys. Rev. {\bf 95}, 577 (1954).
	\bibitem{Chodos} A. Chodos, R. L. Jaffe, K. Johnson et al., Phys. Rev. D {\bf 9}, 3471 (1974).
	\bibitem{Kettner} Ch. Kettner, F. Weber, M. K. Weigel et al., Phys.  Rev.  D {\bf 51}, 1440 (1995).
	\bibitem{Alcock} C. Alcock and A. Olinto, Annu. Rev. Nucl. Part. Sci. {\bf 38}, 161 (1988).
	\bibitem{Peng} G. X. Peng et al., Phys. Rev. C {\bf 62}, 025801 (2000).
	\bibitem{Cleymans} J. Cleymans, R.V. Gavai and E. Suhonen, Phys. Rep. {\bf 130}, 217 (1986).
	\bibitem{Backes} B. C. Backes et al., J. Phys. G: Nucl. Part. Phys. {\bf 48}, 055104 (2021).
	\bibitem{Schwarzschild} K. Schwarzschild, Sitzungsberichte der Koniglich Preussischen Akademie der Wissenschaften Berlin (Mathematical Physics) 189-196 (1916).
	\bibitem{Guver} T. G\"uver et al., Astrophys. J. {\bf 719}, 1807 (2010).
	\bibitem{Guver1} T. G\"uver et al., Astrophys. J. {\bf 712}, 964 (2010).
	\bibitem{Ozel} F. \"Ozel, T. G\"uver and D. Psaltis, Astrophys. J. {\bf 693}, 1775 (2009).
	\bibitem{Rawls} M. L. Rawls et al., Astrophys. J. {\bf 730}, 25 (2011).
	\bibitem{Abubekerov}  M. K. Abubekerov et al., Astron.Rep. {\bf 52}, 379 (2008). 
	\bibitem{Kolassis} C.A. Kolassis, N.O. Santos, D. Tsoubelis, Class. Quantum Grav. {\bf 5}, 1329 (1988)
	\bibitem{Hawking} S.W. Hawking, G.F.R. Ellis, {\it The Large Scale Structure of Spacetime} (Cambridge University Press: Cambridge, UK, 1973).
	\bibitem{Wald} R. Wald, {\it General Relativity} (University of Chicago Press: Chicago, IL, USA, 1984).
	\bibitem{Brassel} B.P. Brassel, S.D. Maharaj, R. Goswami, Entropy {\bf 23}, 1400 (2021)
	\bibitem{Brassel1} B.P. Brassel, S.D. Maharaj, R. Goswami, Prog. Theor. Exp. Phys. {\bf 2021}, 103E01 (20 pages) (2021)
	\bibitem{Gron} \O~Gr\o n , Phys. Rev. D {\bf 31}, 2129 (1985).
	\bibitem{Herrera1} L. Herrera, Phys. Lett. A {\bf 165}, 206 (1992)
	\bibitem{Abreu} H. Abreu, H. Hern{\'a}ndez, L. A. N{\'u}{\~n}ez, Class. Quantum Gravity {\bf 24}, 4631 (2007).
	\bibitem{Pretel} J.M.Z. Pretel, Eur. Phys. J. C {\bf 80}, 726 (2020).
	\bibitem{Chan} R. Chan, L. Herrera, N.O. Santos, Mon. Not. R. Astron. Soc. {\bf 265}, 533 (1993).
\end{thebibliography}
\end{document}